\documentstyle[aps,epsf,epsfig]{revtex}
\def \be   {\begin{equation}}
\def \ee   {\end{equation}}
\def \l {\label}
\begin{document}
\input epsf
\baselineskip=25pt
\title{Dynamics and causality constraints in field theory}
\author{Manoelito M de Souza\footnote{e-mail:manoelit@cce.ufes.br}}
\address{Universidade Federal do Esp\'{\i}rito Santo - Departamento de
F\'{\i}sica\\29065.900 -Vit\'oria-ES-Brasil}
\date{\today}
\maketitle
\begin{abstract}
\noindent We discuss the physical meaning and the geometric interpretation of  causality implementation in classical field theories. Causality is normally implemented through kinematical constraints on fields but we show that in a zero-distance limit they also carry a dynamical information, which calls for a revision of our standard concepts of interacting fields. The origin of infinities and other inconsistencies in field theories is traced to fields defined with support on the lightcone; a finite and consistent field theory requires a lightcone generator as the field support.
\end{abstract}
\begin{center}
PACS numbers: $03.50.De\;\; \;\; 11.30.Cp$
\end{center}

The propagation of a massless field on a flat spacetime of metric $\eta_{\mu\nu}=diag(-1,1,1,1),$ is restricted by
\be
\label{il}
\Delta x^{2}=0,
\ee 
which, with an event $x$ parameterized by  $x^{\mu}=(t,{\vec x})$, defines a a local double (past and futur) lightcone: $\Delta t=\pm|\Delta{\vec x}|.$ This is also a mathematical expression for local causality in the sense that it is a restriction for the massless field to remain on this lightcone, a three-dimensional-hypersurface. It is a particular case of the more general expression
\be
\label{A}
\Delta\tau^2=-\Delta x^{2}.
\ee
which is, besides, the definition of the proper time $\tau.$ For a massive relativistic field there is a restriction on $\Delta\tau$: $$0<|\Delta\tau|\le|\Delta t|$$ which means that the field must remain inside the lightcone (\ref{il}). 
 
The equation (\ref{A}) is just a well-known kinematic restriction on the propagation of a physical object, but it carries a further not implicitly stated dynamical restriction on the interaction between two physical objects.  To highlight this point and discussing some of its consequences is the main objective of this note.\\
Let us consider, for example, $z(\tau)$, the worldline of a classical point electron parameterized by its propertime $\tau,$ and x, an event not in this wordline.  $R=x-z(\tau)$ defines a family of four-vectors connecting the event $x$ to events in the electron worldline $z(\tau).$ The  electromagnetic field in $x,$ of which the charge at $z(\tau)$ is the source, must obey the following kinematic restriction
\be
\l{lc}
R^{2}=0,
\ee
which accounts for the field masslessness.  Then (\ref{lc}) defines a double lightcone with vertex at $x,$ which intercepts $z(\tau)$ at two points: $z(\tau_{ret})$ and $z(\tau_{adv}).$ See  the Figure 1. It is clear that (\ref{lc}) is the same kinematical restriction as (\ref{il}): the retarded field emitted by the electron at $z(\tau_{ret})$ must remain in the $z(\tau_{ret})$-futur-lightcone, which contains x; and according to the standard interpretation \cite{Rohrlich,Jackson,Parrot}, the advanced field produced by the electron at $z(\tau_{adv})$ must remain in the $z(\tau_{adv})$-past-lightcone, which also contains x. All this is just old plain kinematics.\\
\vglue-5cm

\hspace{-1cm}
\parbox[t]{5.0cm}{
\begin{figure}
\vglue-3cm
\epsfxsize=400pt
\epsfbox{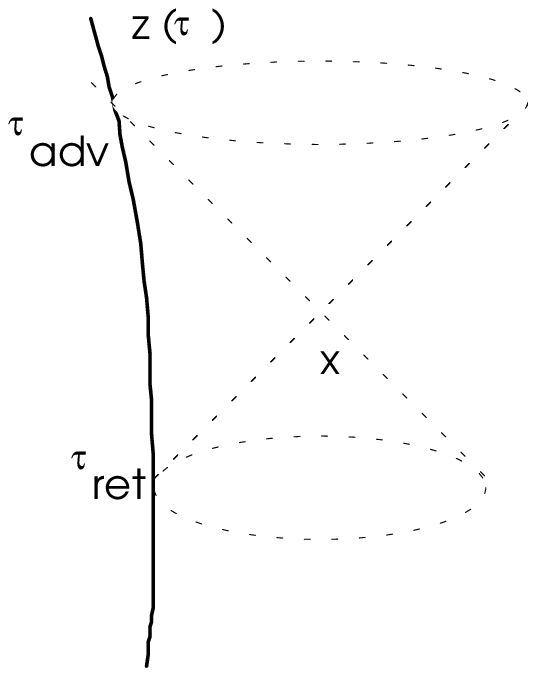}
\vglue-5cm
\end{figure}}\hfill
\\ \mbox{}
\hspace{5cm}
\parbox[t]{8.0cm}{\vglue-5cm 
Fig.1. The advanced and the retarded Li\`enard-Wiechert fields at an event x. $\tau_{adv}$ and $\tau_{ret}$ are the two intersections of the double cone $R^{2}=0$, for $R=x-z(\tau)$, with the electron worldline $z(\tau)$.}\\ \mbox{}
\vglue-3cm

If we now move our attention from $x$ to any other event, let's say, $x+dx$, the lightcones of the constraint (\ref{lc}) will displace their intersections with the electron worldline to $z(\tau_{s}+d\tau)$, where $\tau_{s}$ stands for $\tau_{ret}$ and $\tau_{adv}$, respectively. In other words, we must require that the constraint (\ref{lc}) be continuously applied to all allowed pairs of events ${\big(}x,z(\tau){\big)}$, or, with this assumed hypothesis of continuity, that besides the restriction (\ref{lc}) we must also consider
\be
\l{B}
{\big(}x+dx-z(\tau_{s}+d\tau){\big)}^{2}=0.
\ee
But this corresponds to
\be
\label{C}
R.dR=R.(dx-Vd\tau)=0,
\ee
where $V=\frac{dz}{d\tau}{\Big|}_{\tau_{s}}$ is the electron instantaneous four-velocity at $\tau_{s}$. So we may write
\be
\l{dlc}
d\tau+K.dx=0,
\ee
where K defined by
\be
\l{K}
K=\frac{R}{-V.R}{\Big|}_{\tau_{s}},
\ee
 is a null $(K^{2}=0)$ four-vector, tangent to the lightcone (\ref{lc}). K shows the direction of propagation of the electromagnetic field  emited by the electron at $\tau_{s},$ but as it will be shown in the sequence, it carries further, not so well-known physical informations. The equation (\ref{dlc}), as it can be formally obtained from a derivation of (\ref{lc}),  has wrongly been considered as if all its effects were already described by (\ref{lc}), included in (\ref{lc})  and not, as it is the case, a new an independent restriction to be regarded at a same footing with (\ref{lc}). An evidence of this is that (\ref{lc}) and (\ref{dlc}) carry distinct and complementary physical informations, as we show now. The equation (\ref{lc}) is the same as equation (\ref{il}): it could be no more than a mere particularity, in the above example, that $z(\tau_{s})$ is an event on the electron worldline as (\ref{lc}) is valid for any pair of events on a lightcone generator. Then it, really, would be just an statement that R is a null vector,     a restriction of (\ref{A}) for light-like intervals. Nothing else. There is, however, an implicitly assumed information that the field propagates with the speed of light, but in order to actually convey this further information it is necessary to add to (\ref{lc}) the idea of continuity. Then results (\ref{dlc}), which is, in this context, a consistency relation of (\ref{lc}); it assures that (\ref{lc}) is valid for all successive pair of events $(x,z(\tau)).$ 
Equation (\ref{dlc})  connects restrictions on the propagation of two distinct physical objects, the electron and its field: $d\tau$ describes a displacement of the electron on its worldline while $d x$ is the four-vector separation between two other points where the electron self-field is being considered. If $d\tau=0$ then $dx$ is light-like and colinear to K, as $K.dx=0.$  Thus, $dx$ is  related to a same electromagnetic field at two distinct times. The electromagnetic field at $x+dx$ can be seen as the same field at $x$ that has propagated to there with the speed of light. On the other hand, if $d\tau\ne0$ then $dx$ is not colinear to K and it is related to two distinct electromagnetic signals, emited at distinct times. See figure 2.  

\begin{figure}
\vglue-5.5cm
\hspace{-1.5cm}
\epsfxsize=400pt
\epsfbox{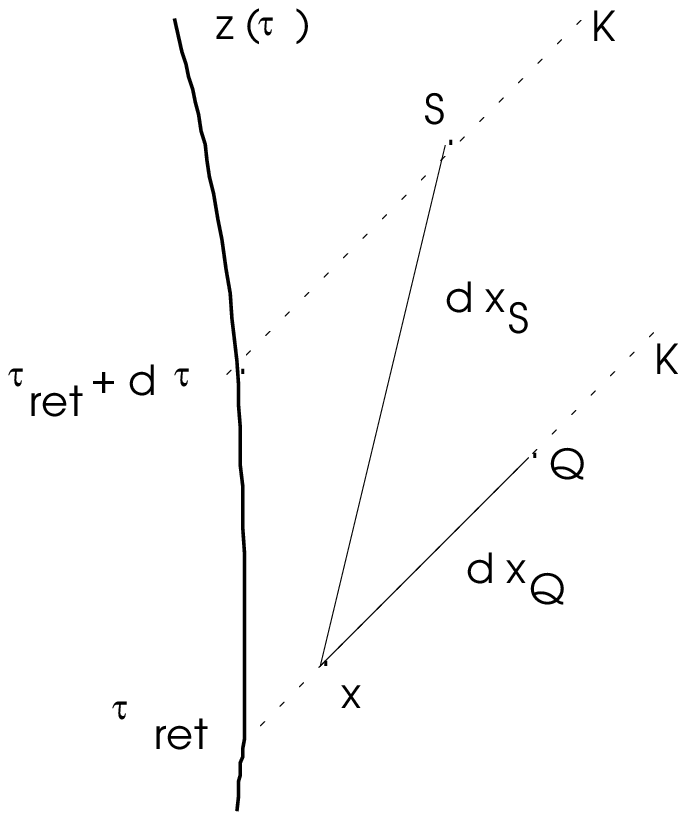}
\end{figure}\hfill
\\ \mbox{}
\hspace{6cm}
\parbox[t]{8.0cm}{\vglue-12cm 
Fig.2. The field at the point Q may be considered as the same field at $x$ that has propagated to Q, because $dx_{Q}$ is colinear to K. The fields at events $x$ and S are two distinct signals emited by the charge at two distinct times $\tau_{s}$ and $\tau_{s}+d\tau\;$ as $\;dx_{S}$ is not colinear to K.}\\ \mbox{}
\vglue-9cm

In this case, the field at $x+dx$ cannot be seen as the same field at $x$ that has propagated to there.\\
Eq.(\ref{dlc}) connects the restriction on the propagation of the charge to the restriction on the propagation of its emitted or absorbed fields. So, it clearly has more information than (\ref{lc}), and not less as if it were just  its derivative. Equations (\ref{lc}) and (\ref{dlc}) are both kinematical relations. In the short distance limit, when $x$ tends to $z(\tau)$, the restriction (\ref{dlc}) is directly related to the changes in the charge movement due to the emission or to the absorption of the electromagnetic field, that is, to the charge and its self-field interaction proccess.  Therefore, in this short distance limit (\ref{dlc}) must  also carry dynamical information, not only kinematical, as in the case of (\ref{lc}). 
What happens to (\ref{lc}) and (\ref{dlc}) in the limit when the event $x$ approaches the event $z(\tau_{s})$?  Nothing obviously happens to (\ref{lc}); $\Delta x$ just goes to zero. To (\ref{dlc}) the restriction connecting $d\tau$ to $dx$ becomes indeterminated because K is not defined in this limit:
\be
\l{i}
\lim_{x\to{z(\tau_{s})}}K=\lim_{x\to{z(\tau_{s})}}\frac{R}{-V.R}=\frac{0}{0}?
\ee 
Our immediate task will be the resolution of this indetermination. First of all it is necessary to precise the definition of this limit: through which path is the event $x$ approaching the event $z(\tau_{s})$? Before answering to that question we must understand the geometric meaning of (\ref{lc}) and (\ref{dlc}). As we have seen, (\ref{lc}) defines a double lightcone and (\ref{dlc}), for $d\tau=0$, defines its tangent hyperplane, actually a family of tangent hyperplanes, labelled by $K^{\mu}=\frac{R^{\mu}}{-V.R}{\Big|}_{\tau_{s}},$ envelloping the lightcone (\ref{lc}). Therefore, for a light-like signal, (\ref{lc}) and (\ref{dlc}) together define a lightcone generator tangent to $K^{\mu}$. The constraint (\ref{lc}) requires that the pair of events $x$ and $z(\tau)$ belongs to a same lightcone while (\ref{dlc}) requires that these events belong to a same tangent hyperplane. Together they imply that $x$ and $z(\tau)$ belong to a same lightcone generator, as described by (\ref{lc}) and (\ref{dlc}) . 
This erases the ambiguity on how $x$ approaches $z(\tau_{s})$ in (\ref{i}), which  can now be written  as
\be
\l{li} 
\lim_{x\to{z(\tau_{s})}}K{\bigg|}_{R^2=0\atop{R.dR=0}}=\frac{0}{0}?
\ee
 This notation intends to denote that $x$ approaches $z(\tau_{s})$ through a K-lightcone generator, that is by the straight line intersection of the cone $(R^{2}=0$) and its tangent hyperplane $(R.dR=0$ or $d\tau+K.dx=0$). This eliminates the ambiguity in the definition of (\ref{i}) at $\tau_{s}$ but does not resolve its indetermination. To resolve it we apply the L'H\^opital's rule evaluating K on the neighboring events of $z(\tau_{s})$ along the electron worldline, i.e., at either $\tau_{s}+ d\tau$ or $\tau_{s}- d\tau.$ This corresponds to replacing the above simple limit of $x\rightarrow z(\tau_{s})$ by a double and simultaneous  limit of $x\rightarrow z(\tau)$ along the K-lightcone generator while $z(\tau)\rightarrow z(\tau_{s})$ along the electron worldline. This simultaneous double limit is pictorially best described by the sequence of points S, Q,...,P
 in the Figure 3; each point in this sequence belongs to a K-generator of a light cone with vertex at the electron worldline $z(\tau).$ 
Then we have for (\ref{i}),
\be
\l{ll}
\lim_{x\to{z(\tau)}\atop\tau\to\tau_{s}}K{\Big|}_{R^2=0\atop{R.dR=0}}=\lim_{x\to{z(\tau)}\atop\tau\to\tau_{s}}\frac{{\dot R}}{-\hbox{\Large a}.R+V.{\dot R})}{\bigg|}_{R^2=0\atop{R.dR=0}}=\lim_{x\to{z(\tau)}\atop\tau\to\tau_{s}}\frac{{-V}}{V.V}{\bigg|}_{R^2=0\atop{R.dR=0}}=V,
\ee

\begin{figure}
\vglue-8cm
\hspace{-1cm}
\epsfxsize=400pt
\epsfbox{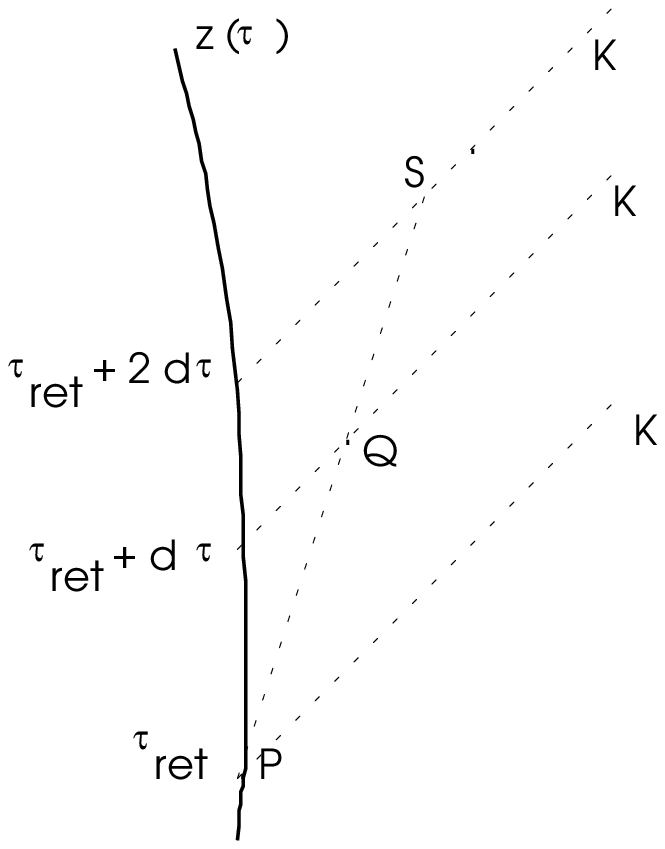}
\end{figure}\hfill
\\ \mbox{}
\hspace{6cm}
\parbox[t]{8.0cm}{\vglue-12cm 
Fig. 3. The double limit $x\rightarrow z(\tau_{ret})$ along the SQ...P line consists of $x\rightarrow z(\tau)$ along the lightcone generator K while $\tau\rightarrow\tau_{ret}$ on the electron worldline.}\\ \mbox{}
\vglue-8cm

as ${\dot R}:=\frac{dR}{d\tau}=-V$ and $V^{2}=-1.$ So $K{\bigg|}_{x=z(\tau_{s})}$ is indefinite but $K{\bigg|}_{x=z(\tau_{s}+)}=V.$  Later on we discuss the physical meaning of this.\\
It is a remarkable and unexpected result: the light-like four-vector K changes into the time-like four-vector V in the above defined (double) limit of $R\rightarrow0;$ it shall change our vision of field theory. \\
Let us consider, as an example, the retarded Lienard-Wiechert solution of Classical Electrodynamics \cite{Rohrlich,Jackson,Parrot,Teitelboim,Rowe,Lozada}
\be
\l{aA}
A^{\mu}(x)=\frac{eV^{\mu}(\tau)}{\rho}{\bigg|}_{\tau=\tau_{ret}},\qquad\hbox{ for}\quad \rho\ne0,
\ee
where $\tau_{ret}$ is the retarded solution of the constraint (\ref{lc}),  imposed to $A(x),$  and  $\rho=-V.R$ represents $|{\vec R}|$ in the charge rest-frame.  Observe that $A(x)$ is restricted just by (\ref{lc}) and so its support is the lightcone, but for calculating its Maxwell field, ${F_{\mu\nu}}:=\nabla_{\nu}A_{\mu}-\nabla_{\mu}A_{\nu}$, it is necessary to consider also the constraint (\ref{dlc}). The field $F$, in contradistinction to the field A, is then constrained by (\ref{lc}) and by (\ref{dlc}) and has, therefore, a lightcone generator K as a support manifold. $F$ is then erroneously taken as defined , like A, on a lightcone, but actually it should carry a label K, as in $F_{K}$, showing its restricted manifold support. The failure on understanding this, as we shall see now, is the cause of many confusions and misconceptions in Classical Electrodynamics. The constraint (\ref{dlc}) implies on
 \be
K_{\mu}=-\frac{\partial\tau_{ret}}{\partial x^{\mu}},
\ee
and then, 
\begin{equation}
\label{nablaA}
\frac{1}{e}\nabla_{\mu}A^{\nu}{\bigg|}_{\tau_{ret}}=\bigg{\{}\nabla_{\mu}\frac{V^{\nu}}{\rho}\bigg{\}}{\bigg|}_{\tau_{ret}}=-\bigg{\{}\frac{K_{\mu}\hbox{\Large
a}^{\nu}}{\rho}+\frac{V^{\nu}}{\rho^{2}}\nabla_{\mu}\rho\bigg{\}}{\bigg|}_{\tau_{ret}}=-\frac{1}{\rho^{2}}{\bigg(}K_{\mu}W^{\nu}+V_{\mu}V^{\nu}{\bigg)}{\bigg|}_{\tau_{ret}},
\end{equation}
with 
\begin{equation}
\label{W}  
W^{\mu}=\rho\hbox{\Large a}^{\mu}+V^{\mu}{\bigg(}1+\rho\hbox{\Large a}.{K}{\bigg)}{\bigg|}_{\tau_{ret}},
\end{equation}
$\hbox{\Large a}:=\frac{dV}{d\tau},$ and
\be
\nabla_{\mu}\rho{\bigg|}_{\tau_{ret}}={\bigg\{}K_{\mu}{\bigg(}1+\rho \hbox{\Large a}.{K}{\bigg)}-V_{\mu}{\bigg\}}{\bigg|}_{\tau_{ret}}.
\ee
Observe that $F$ is indeed explicitly dependent on K,  but due to the continuous differentiability of the lightcone (out of its vertex) there is no problem if we take the lightcone as the support manifold of $F$, as long as we remain out of its vertex.\\
So, we have
\be
\label{F} 
F^{\mu\nu}=\frac{1}{\rho^{2}}{\bigg(}K^{\mu}W^{\nu}-K^{\nu}W^{\mu}{\bigg)}{\bigg|}_{\tau_{ret}},
\end{equation}
The electron self-field energy-momentum tensor, $4\pi\Theta=F.F-\frac{\eta}{4}F^{2}$, is then
\begin{equation}
\label{t'}
-4\pi\rho^{4}\Theta=(KW+WK)+KKW^{2}+WWK^{2}+\frac{\eta}{2}(1-K^{2}W^{2}), 
\end{equation}
as $K.V=-1$ from (\ref{K}) and $K.W=-1$. The expressions (\ref{aA}-\ref{t'}) are all constrained by (\ref{lc}), i.e. $\tau=\tau_{ret},$ and are all only valid for $\rho\ne0$, region where $K^{2}=0.$ So, instead of (\ref{t'})  we may write 
\be
\l{rt}
-4\pi\rho^{4}\Theta{\bigg|}_{K^{2}=0}=(KW+WK)+KKW^{2}+\frac{\eta}{2},\qquad\hbox{ for}\quad \rho>0,\;\tau=\tau_{ret},
\ee
which corresponds to the usual expressions that one finds, for example in \cite{Rohrlich,Jackson,Parrot,Teitelboim,Rowe,Lozada}. They are equivalent, as long as $\rho>0$.
But there is a well-known major problem in Classical Electrodynamics. The four-vector momentum associated to the electron self-field is defined by the flux of its $\Theta$ through a hypersurface $\sigma$ of normal n:
\be
\l{P}
P=-\int d^{3}\sigma\Theta.n{\bigg|}_{K^{2}=0},
\ee
but  $\Theta$ contains a factor $\frac{1}{(\rho)^{4}}$ and this would make P highly singular at $\rho=0$, that is at $x=z(\tau_{ret}),$ if P were defined there!  This is the old well-known self-energy problem of Classical Electrodynamics which heralds \cite{JS} similar problems in its quantum version. This divergence at $\rho=0$ is also the origin of nagging problems on finding a classical equation of motion for the electron \cite{Rohrlich,Jackson,Parrot,Teitelboim,Rowe,Lozada}.\\
Previous attempts, based on distribution theory, for  taming this singularity have relied on modifications of the Maxwell
theory with  addition of extra terms to $\Theta{\Big|}_{K^{2}=0}$ on the electron world-line (see for example the reviews \cite{Teitelboim,Rowe,Lozada}). They redefine $\Theta{\Big|}_{K^{2}=0}$ with the inclusion of extra terms that are non-null only at the electron world-line; this makes $\Theta$ integrable without changing it at $\rho>0,$ and so preserving the standard results of Classical Electrodynamics. But this is always an ad hoc introduction of something strange to the theory. Another unsatisfactory aspect of this procedure is that it regularizes the above integral but leaves an unexplained and unphysical discontinuity in the flux of four-momentum, 
\be
\l{pp}
\int dx^{4}\Theta^{\mu\nu}\nabla_{\nu}\rho\;\delta(\rho-\varepsilon_{1}),
\ee
 through a  cylindrical hypersurface $\rho=\varepsilon_{1}\sim0$ enclosing  the charge world-line.

It is clear now, after equation (\ref{ll}), that the standard practice of replacing $\Theta$ by $\Theta{\bigg|}_{K^{2}=0}$ is not justified and, more than that, it is the cause of the above divergence problem and the related misconceptions in Classical Electrodynamics. We must use (\ref{t'}), the complete expression of $\Theta$, in (\ref{P}) and repeat for it the same double limit done in (\ref{ll}). The long but complete and explicit calculation is done in \cite{hep-th/9610028}; here we just give the results:\\ $P{\bigg|}_{x=z(\tau_{s})}$ is undefined but
\be
\l{PO}
P{\bigg|}_{x=z(\tau_{s}-)}=0;\quad P{\bigg|}_{x=z(\tau_{s}+)}=0.
\ee
There is no infinity at $\rho=0$! It is amusing that for avoiding the infinity of (\ref{P}) at $\rho=0$ instead of adding anything  to $\Theta$ one just should not drop anything from it! This infinity disapears only when the double limiting proccess is taken because then the lightcone generator $K$ must be recognized as the actual support of the Maxwell field $F.$ The infinities and other inconsistencies of Classical Electrodynamics are not to be blamed on the point electron but on the lightcone support of (\ref{aA}).\\
Observe that the discontinuity pointed in (\ref{pp}) for $\rho=\varepsilon_{1}\sim0$ still remains because for $\rho>0$, that is out of the double limiting, prevails the lightcone support of (\ref{aA}). In order to be consistent Classical Electrodynamics must be entirely formulated in terms of fields with support on the lightcone generators, not on the lightcone. This  has been done in references \cite{hep-th/9610028}, \cite{hep-th/97xxxxx} and \cite{dce}. It is also shown in reference \cite{hep-th/9610028} that the infinities associated to the electron bound momentum  and to the Schot term in the Lorentz-Dirac equation disapear too; the Lorentz-Dirac equation is not correct.  Actually there can be no electron equation of motion in Classical Electrodynamics but only an effective equation based on average field values. The point is that the Maxwell-Faraday concept of a field with support on the lightcone, like (\ref{aA}), cannot represent the fundamental field (a single photon). Like the Maxwell stress tensor $F$ in (\ref{F}) the classical representation of the fundamental field has to be defined on a lightcone generator.\\
Equation (\ref{PO}) shows that $z(\tau_{ret})$ is an isolated singularity. This is in direct contradiction to the standard view of a continuous field, emited or absorbed by the charge in a continuous way. According to (\ref{PO}) there is no charge self field at $z(\tau_{s}\pm d\tau)$, but only sharply at $z(\tau_{s})$.
This is puzzling! It is saying that the Gauss' law, in the zero distance limit, $lim_{S\rightarrow0}\int_{S}d\sigma{\vec E}.{\vec n} = 4\pi  e$, is valid only at $z(\tau_{ret})$ and not at $z(\tau_{ret\pm})$ because ${\vec E}(\tau_{ret})\ne0$ but ${\vec E}(\tau_{ret\pm d\tau})=0.$\\
It implies, in other words, that the electromagnetic interactions are discrete and localized in time and in space. In terms of a discrete field interaction along a lightcone generator, as the one represented in the Figure 4, we can understand the physical meaning of (\ref{li}), (\ref{ll}) and of (\ref{PO}). 
The Maxwell fields are just effective average discriptions of an actually discrete interaction field. The field discreteness (or the existence of photons) is masqueraded by this averaged field and it takes the zero distance limit to be revealed from the Maxwell field. This could have been taken, if it were known at the begining of the century, as a first indication of the quantum, or discrete nature of the electromagnetic interaction. All these are consequences of the dynamical constraints hidden on the restrictions (\ref{il}) and (\ref{dlc}).

\begin{figure}
\vglue-3.5cm
\epsfxsize=400pt
\epsfbox{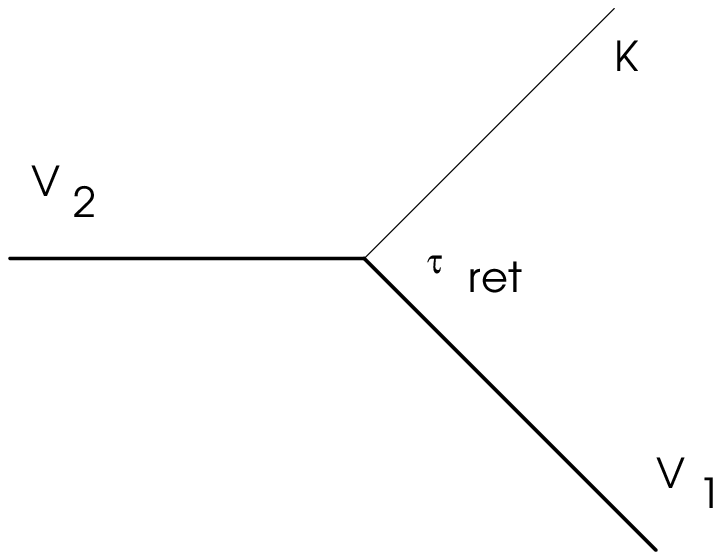}
\vglue-5cm
\end{figure}\hfill
\\ \mbox{}
\hspace{7cm}
\parbox[t]{8.0cm}{\vglue-13.5cm 
Fig.4. A discrete interaction along a lightcone generator $K.$ There is no electron self-field before or after $\tau_{ret}.$  This is an isolated singular point on the electron worldline only because its tangent is not defined there; there is no infinity.}\\ \mbox{}
\vglue-12cm

\end{document}